\documentclass[%
 reprint,
superscriptaddress,
 amsmath,amssymb,
 aps,prapplied,floatfix
]{revtex4-2}

\usepackage{graphicx}
\usepackage{dcolumn}
\usepackage{bm}
\usepackage{hyperref}
\hypersetup{pdfstartview={FitH},pdfpagemode={UseNone}, colorlinks=true,bookmarks=false,allcolors=blue, breaklinks=true, bookmarksopen=true, pdfnewwindow=true}
\usepackage[all]{hypcap}

\graphicspath{{figures/}{}}

\usepackage[english]{babel}

\begin{document}


\title{Optical Neural Network Based on Synthetic Nonlinear Photonic Lattices}

\author{Artem V. Pankov}
\author{Ilya D. Vatnik}%
\email{i.vatnik@nsu.ru}
\affiliation{%
Novosibirsk State University, Pirogova str. 2, Novosibirsk 630090, Russia
}%

\author{Andrey A. Sukhorukov}
\email{andrey.sukhorukov@anu.edu.au}
\affiliation{
Research School of Physics, Australian National University, Canberra, ACT 2601, Australia
}
\affiliation{
ARC Centre of Excellence for Transformative Meta-Optical Systems (TMOS), Australia
}

\date{\today}

\begin{abstract}
We reveal that a synthetic photonic lattice based on coupled optical loops can be utilized as a neural network for processing of optical pulse sequences in time domain. As a proof-of-concept, we train the optical system to restore an initial shape of the pulse train from the signal distorted due to linear dispersion in a fiber-optic link. We also show efficient training of the optical network with an intrinsic Kerr-type nonlinearity for the realization of target nonlinear transmission functions and inference functionality for the discrimination of different pulse sequences.
The theoretical modeling is performed under practical conditions and can guide future experimental realizations.
\end{abstract}

\maketitle


\section{Introduction}
Photonic systems are actively developed for the realization of neuromorhpic optical computing~\cite{Shastri:2021-102:NPHOT}. A paradigm of neural networks processing distributed information is rapidly advancing from computational models to the physical optical implementations. Such optical neural networks (ONNs) process data through parallel light redistribution in space, time, or frequency domains. 
Due to a much higher bandwidth, ONNs can overcome their purely electronic counterparts in both the processing speed and energy efficiency~\cite{Nahmias:2020-7701518:ISQE}.  Scalable realizations of interconnections performing massive amounts of linear operations makes ONNs promising in plenty of applications, spanning from the nonlinear optimization problems~\cite{deLima:2019-1515:JLT} to signal processing~\cite{Zhang:2019-3033:NCOM}. 

Photonic neural networks were extensively developed for operations with spatially- and frequency-encoded optical data~\cite{Shen:2017-441:NPHOT, Tait:2014-4029:JLT, Tait:2019-64043:PRAP, Feldmann:2019-208:NAT, Shainline:2017-34013:PRAP, Xu:2021-44:NAT, Feldmann:2021-55:NAT}. On the other hand, the ONN realization in time domain with coherent manipulation of optical pulse trains could lead to advances in communications and real-time data analysis~\cite{Huang::2020}. There are complementary developments of  
photonic reservoir computing in time domain~\cite{Vandoorne:2014-3541:NCOM, Tanaka:2019-100:NNW, Chembo:2020-13111:CHA}, 
however such systems tend to suffer from high latency associated with post-processing involving electronic components.

Incorporating a nonlinear response of neurons into an optical neural network is one of the main challenges that still limits the development of such systems. 
In experimental realizations of ONNs, the nonlinear transmission function was commonly implemented through 
optical-electrical-optical conversion of the signal~\cite{Nahmias:2016-151106:APL, Williamson:2020-7700412:ISQE, McCaughan:2019-451:NELT}, which restricts the performance and scope of potential applications.

We note that there is a remarkable system called synthetic photonic lattice (SPL) that is developed to operate with optical pulse trains in controllable manner~\cite{Schreiber:2010-50502:PRL, Regensburger:2011-233902:PRL, Regensburger:2012-167:NAT}. The most common fiber-optical realization of the device based on a pair of coupled recirculating fiber loops was demonstrated as a versatile tool for the investigation of various physical processes in synthetic dimensions, including topological effects~\cite{Wimmer:2017-545:NPHYS, Chen:2018-100502:PRL, Pankov:2019-3464:SRP, Weidemann:2020-311:SCI, Bisianov:2020-53511:PRA}. Furthermore, such platforms can incorporate nonlinear optical fibers, 
enabling the observation of soliton formation and other nonlinear effects~\cite{Wimmer:2013-780:NPHYS, Wimmer:2015-7782:NCOM}.

In this work, we propose a synthetic photonic lattice to serve as an efficient ONN platform for processing optical pulse trains, both in the linear regime and 
involving optical nonlinear response. The pulses propagating through the SPL multiply and interfere with each other at different times, effectively representing signals coming to and from different neurons.
We theoretically demonstrate the general applicability of the optical neural network based on SPL by considering three representative tasks. First, we illustrate the restoration of a pulse series distorted by a linear dispersion in a communication link.
Second, we reveal that tailored nonlinear transmission functions, involving the amplitude and phase modulation, can be realized by an SPL based on the Kerr optical nonlinearity through the training of linear elements.
Third, we show that a nonlinear SPL trained for inference can efficiently discriminate a set of non-orthogonal input states, such that each input is mapped to one specific output slot.

\begin{figure}[th]
\centering
\includegraphics[width=\linewidth]{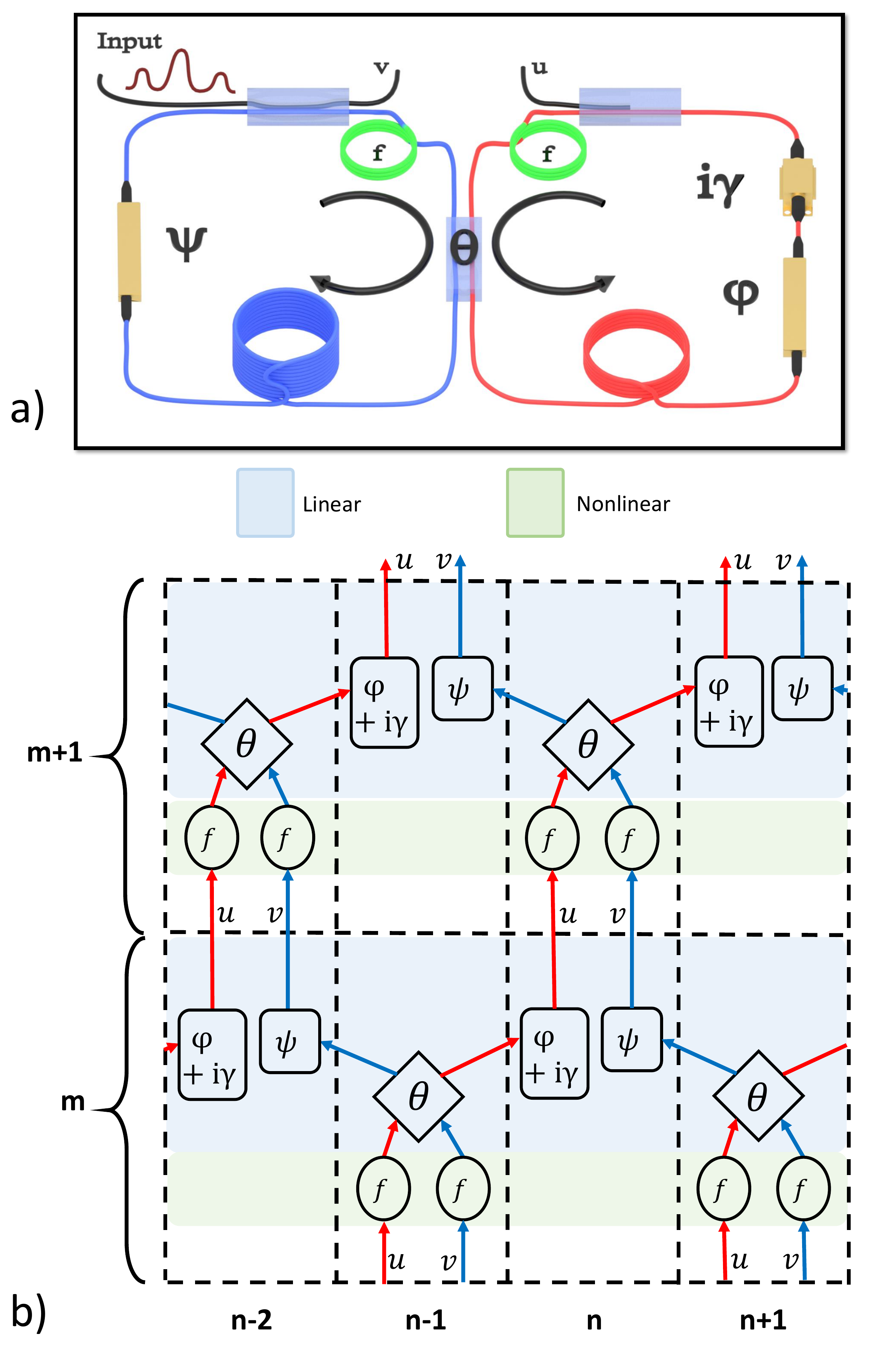}
\caption{(a)~A synthetic photonic lattice with controllable coupling coefficients $\theta$, linear phases ($\phi$,$\psi$), gain $i\gamma$, and nonlinear elements $f$.  (b)~Equivalent optical neural network with controllable synapse weights and nonlinear transfer function operating in 2D synthetic space.}
\label{fig:scheme}
\end{figure}

\section{Synthetic photonic lattice as optical neural network}

We analyze the dynamics of optical pulses in a synthetic photonic lattice and reveal that this platform can realize a reconfigurable optical neural network, suitable for practical implementations using, for instance, already demonstrated fiber-optical experimental setups~\cite{Muniz:2019-6013:OL}. 

\subsection{Synthetic photonic lattice}

We consider a synthetic photonic lattice consisting of two optical loops 
connected via a tunable directional coupler ($\theta$) as schematically illustrated in Fig.~\ref{fig:scheme}(a), similarly to the previous experimental fiber-optical implementations~\cite{Muniz:2019-6013:OL}.
The length of loops is specially made different, such that the pulses propagating in the long loop are delayed with respect to pulses in the short loop by $\Delta \tau=T_{\rm long}-T_{\rm short}$. As pulses can jump between the loops, this results in coupling between the pulses from different time slots, delayed by integer multiples of $\Delta \tau$. For example, a single pulse launched into the system through an additional coupler produces a pulse train, representing a photonic quantum walk or optical discrete diffraction~\cite{Schreiber:2010-50502:PRL}. Accordingly, the complex pulse amplitudes effectively evolve in a synthetic 2D discrete space indexed by a pair of integer indices $(m,n)$. Here $m$ is a mean number of roundtrips with a characteristic time $T_{\rm rt}=(T_{\rm long}+T_{\rm short})/2$, and $n \Delta \tau$ is a time delay of a particular pulse within a roundtrip. To avoid the overlap of pulse trains from different roundtrips, a condition 
\begin{equation} \label{eq:Trt}
   |2N \Delta \tau| < T_{\rm rt}    
\end{equation}
can be ensured in experiments \cite{Schreiber:2010-50502:PRL, Regensburger:2011-233902:PRL}, where $2N$ is the total number of slots. 

The modulators in each loop can be programmed to modify the phase ($\phi$,$\psi$) of individual pulses, whose energy can be also controlled with an optical amplifier with gain $\gamma$. 
We note that the modulation of coupling, phases, and gain parameters can be pre-programmed in the electronic components after an offline network training, thereby allowing for fast optical pulse transformations without a need for electronic read-outs at the intermediate steps.


The evolution of the complex pulse amplitudes in the short ($u_n^m$) and long ($v_n^m$) loops can be modelled by the coupled equations~\cite{Regensburger:2011-233902:PRL, Wimmer:2013-780:NPHYS}
\begin{equation}
\begin{split}
\label{eq:SPL_evolution}
    u_n^{m+1} = &\exp(i\phi^m_n+\gamma^m_n)[  f(|u^m_{n-1}|^2)u^m_{n-1}\cos(\theta^m_{n-1})+\\
    & \qquad i f(|v^m_{n-1}|^2)v^m_{n-1}\sin(\theta^m_{n-1})] \, , \\
    v_n^{m+1} = &\exp(i\psi^m_n)[  f(|v^m_{n+1}|^2)v^m_{n+1}\cos(\theta^m_{n+1})+\\
    & \qquad i f(|u^m_{n+1}|^2)u^m_{n+1}\sin(\theta^m_{n+1})] \, ,
\end{split}    
\end{equation}
where $\phi_n^m$ and $\psi_n^m$ are the controllable phase shifts introduced in the short and the long loops, $\gamma_n^m$ is a gain coefficient, $\theta_n^m$ defines the variable optical coupling between the two loops. The function $f(|x|^2)$ represents the effect of optical nonlinearity. In fiber-optical implementations of the SPL \cite{Wimmer:2013-780:NPHYS, Wimmer:2015-7782:NCOM, Bisianov:2019-63830:PRA}, the Kerr nonlinearity is associated with an intensity-dependent phase shift, 
\begin{equation} \label{eq:Kerr}
    f(|x|^2) = \exp(i \alpha |x|^2 ) , 
\end{equation}
where $\alpha$ is a nonlinear coefficient. The initial state of the system $(u_n^0,v_n^0)$ is determined by the amplitudes of the pulses composing the pulse train  launched into the system through the input coupler, as illustrated in Fig.~\ref{fig:scheme}(a). 
We note that according to Eq.~\eqref{eq:SPL_evolution}, the signals $u^m_{n}$ within the even and odd slots $n$ at the same roundtrip $m$ never couple with each other. Thus, a synthetic photonic lattice with $2 N$ slots consists of two unconnected sublattices of $N$ slots each.

We make an observation that the light evolution in the synthetic 2D space in a single sublattice  effectively represents a structure of a feed-forward deep optical neural network (ONN), as shown in Fig.~\ref{fig:scheme}(b).
The parallelism in such an ONN is realized through time domain multiplexing,
where a single layer corresponds to a roundtrip through the loops, 
while each neuron is the pulse interference at the coupler. Thus the number of hidden layers is proportional to the total number of roundtrips in the SPL, and the number of neurons is equal to the number of pulses within both loops.
Every neuron of the SPL based network has exactly two synapses, which is analogous to ONN implementations with spatially-multiplexed photonic circuits~\cite{Shen:2017-441:NPHOT}. The weights of synapses are determined by the controllable phase shifts $\phi$ and $\psi$, amplification coefficient $\gamma$, and also by the coupling coefficient $\theta$.

Importantly, the nonlinear response in the system can be achieved all-optically through the Kerr-induced phase shift in the fiber spools, as formulated in Eq.~(\ref{eq:SPL_evolution}). Whereas such nonlinearity has the form that drastically differs from the standard rectifier activation utilized in artificial neural networks~\cite{LeCun:2015-436:NAT}, it can enable ultrafast optical processing~\cite{Hughes:2019-eaay6946:SCA}. Furthermore, such nonlinearity was found to be efficient in nonlinear signal equalizers based on electronic neural networks \cite{Hager2018, Sidelnikov:2021-2397:JLT}. 

Overall, the processing time with SPL is $M T_{\rm rt}$, where $M$ is the total number of layers in  
the ONN. As the SPL has each node connected with only two of the subsequent nodes, it is necessary to implement a large enough number of roundtrips to make the optical neural network fully connected. Thus $M=2N+1$ is the minimal necessary number of roundtrips to make all of the input slots to be connected with all of the output slots. Then, the condition in Eq.~\eqref{eq:Trt} sets a lower bound for $T_{\rm rt}$ that is proportional to a number of slots $N$. Typical pulse durations and roundtrip times exploited in state-of-the-art fiber-optics based SPL are $\Delta \tau \sim 100$ ns and $T_{\rm rt} \sim 20$ $ \mu s$ \cite{Weidemann:2020-311:SCI}. In these systems, up to $M=200$ roundtrips were achieved~\cite{Muniz:2019-6013:OL}, which is sufficient to create fully connected ONN with $N \le 100$.

There is a practical potential to design SPL aimed at processing of sequences of shorter pulses with the lower latency. For example, a single pulse duration of 30~ps yields $\Delta \tau \sim 30$~ps and thus $\Delta L=\Delta \tau c\sim 6$~mm.  Such systems can be implemented using electro-optical modulators integrated with lithium niobate waveguides~\cite{Mourgias-Alexandris:2019-9620:OE}, as losses in such systems can be small enough to allow for at least tens of roundtrips in the SPL before the signal is attenuated considerably~\cite{Qi:2020-1287:NANP}. The optical nonlinear response can also be accessed based on Kerr effect or gain saturation \cite{Mourgias-Alexandris:2019-9620:OE}.






\begin{figure*}[t]
\centering
\includegraphics[width=\textwidth]{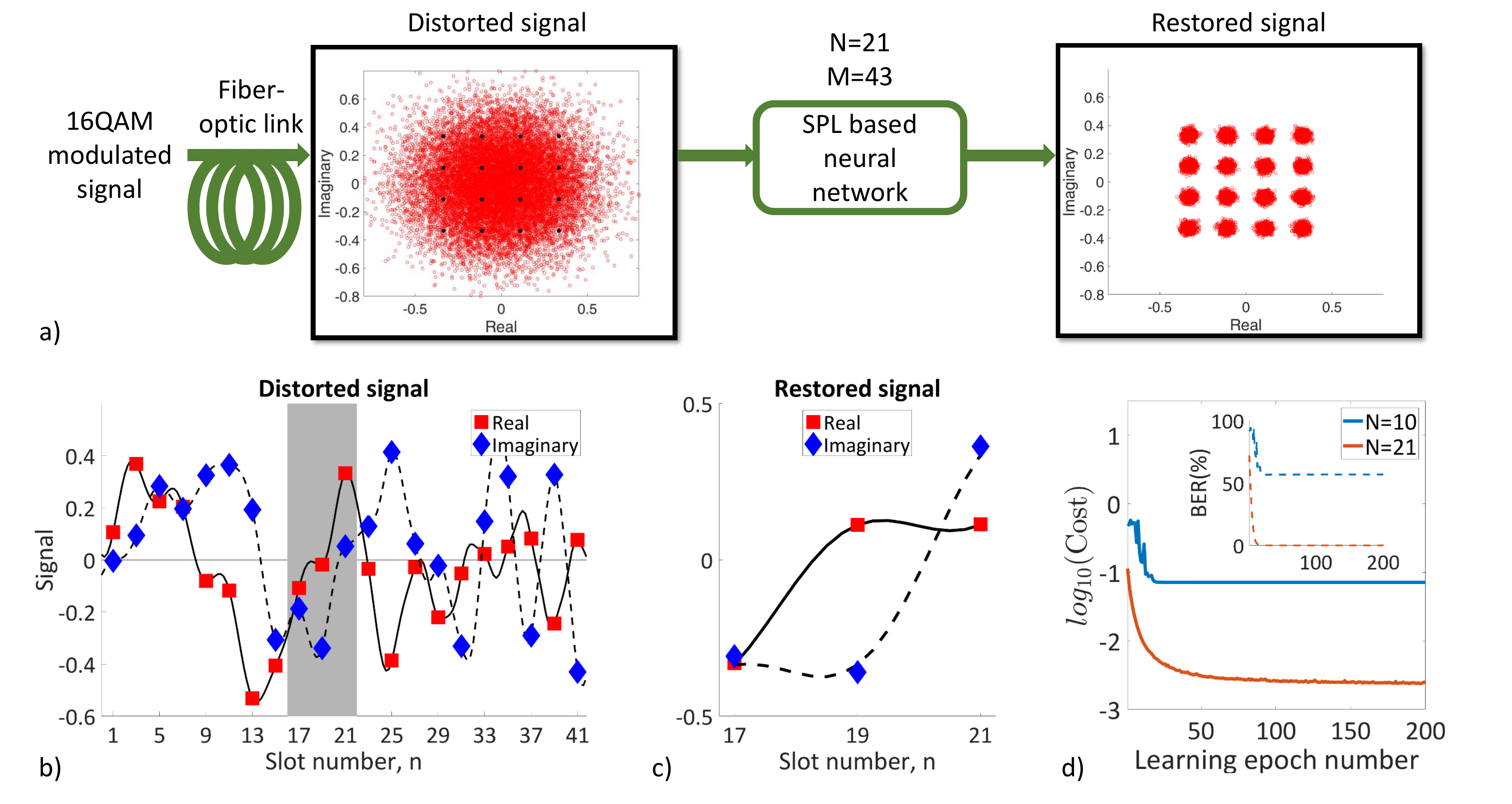}
\caption{Signal restoration after dispersion-induced distortion of 16QAM telecommunication signal with the ONN. (a)~Constellation diagrams for distorted signal and restored with the SPL based neural network. (b)~An example of a distorted signal, with the solid and dashed lines showing the real and imaginary parts. The symbols denote the amplitudes at discrete time steps, which are used as the SPL inputs for a sublattice with odd slot numbers.
(c)~Example of the restored signal at the SPL output: red and blue diamonds show the real and imaginary parts. The solid and dash lines show the real and imaginary parts of the initial undistorted signal for reference. 
(d)~Cost function and bit error rate (BER) during training of the SPL with the different sublattice widths $N=10,\,21$ as indicated by labels.
Results in (a),(c) are shown for $N=21$.
}
\label{fig:telecom}
\end{figure*}

\subsection{Training of the optical network}

We perform offline training of the optical network on a computer. We adopt a commonly used algorithm of stochastic gradient descent through backpropagation~\cite{Hughes:2018-864:OPT} to train the physical parameters of the SPL. A cost function to be minimized is defined as:
\begin{equation}
\label{eq:cost_function}
     C(\textbf{A}^M,\textbf{A}_{target})=\frac
     {\sum\limits_{i=1}^{4N}(V_{shape})_i \left|A^M_i-(A_{target})_i \right|^2}
     {\sum\limits_{i=1}^{4N}(V_{shape})_i \left|(A_{target})_i \right|^2}   
\end{equation}
where $\textbf{A}^m=[\textbf{u}^m,\textbf{v}^m]$ is a combined vector of signals in the short $u_n^m$ and long $v_n^m$ loops at the roundtrip $m$,
$M$ is the total number of roundtrips, and $\textbf{A}_{target}$ is the target output vector. The  $V_{shape}$ is a $4 N$-long vector with the element values of 0 and 1 controlling which output slots are included in the training process.

To minimize the cost function, the parameters of SPL are tuned in accordance to the function gradient. The corrections of SPL coefficients are calculated with standard approach~\cite{Hughes:2018-864:OPT}, adjusted due to the complex-valued nature of optical signals:
\begin{equation}
\begin{split}
\label{eq:parameter_correction}
\Delta x^m_n & = -a\frac{\partial C(\textbf{A}^M,\textbf{A}_{target})}{\partial x_n^m} \\ 
& = -a\sum\limits_{i=1}^{4N}(V_{shape})_i
\left[
\frac{\partial C}{\partial A_{i}^{M}}
\frac{\partial A_{i}^{M}}{\partial x_{n}^{m}} 
+
\frac{\partial C}{\partial {A_{i}^{M}}^{\ast}}
\frac{\partial {A_{i}^{M}}^{\ast}}{\partial x_{n}^{m}}
\right] \, ,
\end{split}
\end{equation}
where $x_n^m=\phi_n^m,\psi_n^m,\theta_n^m$ and $\gamma_n^m$ are the SPL coefficients, 
and $a$ is the learning rate. 
Analytical expressions for the derivatives ${\partial A_{i}^{M}}/{\partial x_{n}^{m}}$ were found from the evolution, Eqs.~(\ref{eq:SPL_evolution}). 

We perform training using a batch stochastic descent approach~\cite{LeCun:2015-436:NAT}. At each epoch, the whole set of training data is randomly divided into batches of a predetermined size and batches-averaged parameters corrections are consequently applied to the ONN. As a starting condition of SPL at the beginning of training procedure, all of the phase modulations and amplifications are set to zero ($\phi=\psi=\gamma=0$), and all of the couplers are set to 50/50 state ($\theta=\pi/4$). At the edges of the synthetic lattice, we implement reflecting boundary conditions by setting $\theta_1^m = \theta_{2N}^m \equiv \pi/2$, and these values are kept fixed during training.



\section{Applications of SPL based neural network}

\subsection{Signal restoration in telecommunication fiber links}
As the proposed ONN operates directly with time sequence of optical signals in fiber, it's natural to consider a potential for applications in optical telecommunication signal processing. Indeed, while mitigation of transmission impairments can be done with digital neural network processing \cite{Sidelnikov:2021-2397:JLT}, all-optical implementations 
may offers advantages in performance \cite{Argyris:2018-8487:SRP}. 

We demonstrate the feasibility of the system for the task of signal equalization.
Specifically, we trained an SPL based optical neural network to restore the signal after distortions introduced due to the dispersion effects accumulated during the propagation through a 100 km fiber span, see a conceptual diagram in Fig.~\ref{fig:telecom}(a). We simulated the propagation of a modulated time-distributed input signal according to 16QAM scheme~\cite{Essiambre:2010:JLT}. The pulse sequences contained up to $2^{14}$ pulses, each having a sinc shape with duration of $30$~ps and carrier wavelength of 1550~nm. Propagation through a standard SMF-28 fiber with the dispersion coefficient of $17$~ps/nm/km was emulated with linear Schr\"odinger equation. Dispersion results in a considerable broadening of each pulse, such that 98\% of the initial power spreads out across 18 neighboring pulse slots. The sampled complex amplitudes for a stream of randomly chosen signals are visualized on a constellation diagram shown in Fig.~\ref{fig:telecom}(a), where the black dots represent the signals before propagation, and red~-- after propagation-induced distortions.

We propose a restoration process where the SPL input comprises a part of the whole output sequence as an input dataset, and then the SPL processes it to restore the initial amplitudes and phases of several central pulses. 
Due to a finite spectral bandwidth of sinc-shaped input pulses, the optical pulse train is fully defined by discretely sampled points as illustrated in Fig.~\ref{fig:telecom}(b). 
The discrete complex amplitudes, determined after simulated propagation through the fiber link, were then processed with the SPL.
The samples were cut into sets of 21 complex amplitudes each to form bunches of input datasets. The central three amplitudes of each twenty-one-size piece, as indicated by shading in Fig.~\ref{fig:telecom}(b), were used to create a single target set for training of the ONN, and accordingly the $V_{shape}$ used in the cost function, Eq.~\eqref{eq:cost_function}, has only three nonzero elements. 

Since dispersion acts as a linear transformation, we consider a linear SPL with $\alpha=0$ in Eq.~\eqref{eq:Kerr}, such that $f(|x^2|) \equiv 1$, and with energy conservation ($\gamma=0$) as a sufficient ONN for the task.
To ensure that the signal is occupying only one of the sublattices of the SPL, we match the slot duration of the SPL to half of the symbol duration of the telecommunication signal. 




We trained the SPL using 
bunches of distorted and target (initial) sequences of pulse amplitudes. 
We used an SPL with $M=2N+1$ to form a fully connected ONN as discussed above.
Training process converges smoothly, and processing of a test dataset with the neural network gives excellent restoration of the initial sequence, see Fig.~\ref{fig:telecom}(a, right),
including both the real and imaginary parts as shown in Fig.~\ref{fig:telecom}(c). 
We tested how wide the system must be to solve the task, and it appeared that the width of SPL sublattice of $N=21$ slots is sufficient as indicated by cost function convergence in Fig.~\ref{fig:telecom}(d), whereas $N=10$ slots is not enough. This is consistent with our estimates of the signal spreading over its 18 neighbours during propagation in a fiber-optical link.
Remarkably, just three epochs of training are sufficient to obtain an SPL that makes the bit error rate (BER), defined as a part of the misidentified symbols in the constellation diagram, equal to zero [Fig.~\ref{fig:telecom}(d)].



\subsection{Nonlinear transmission functions}

The incorporation of nonlinear activation in artificial neural networks is essential for their functionality, similar to the biological counterparts. Yet in many implementations of optical neuromorphic systems, the nonlinearity was realised through electronic detection~\cite{Shen:2017-441:NPHOT, Hamerly:2019-21032:PRX, Xu:2021-44:NAT}. There is active research on developing activation response based on optical nonlinearity~\cite{Sui:2020-70773:IACC, Jha:2020-4819:OL}, which relatively weak effects need to be specially enhanced.



We reveal that SPL can be trained to achieve a desired nonlinear activation response, unconstrained by the type of optical nonlinearity at the microscopic level. We consider the Kerr nonlinearity in optical fibers, which produces an additional phase shift per roundtrip that is proportional to the signal intensity,
and choose the amplitude normalization such that the self-phase-modulation coefficient is $\alpha=1$ in Eq.~\eqref{eq:Kerr}.  
We then aim to train the network to achieve a target nonlinear transformation 
of the signal amplitude in a distinct input time slot $n$ into the same output slot $n$ after a specific number of roundtrips. 

\begin{figure}[ht]
\centering
\includegraphics[width=\linewidth]{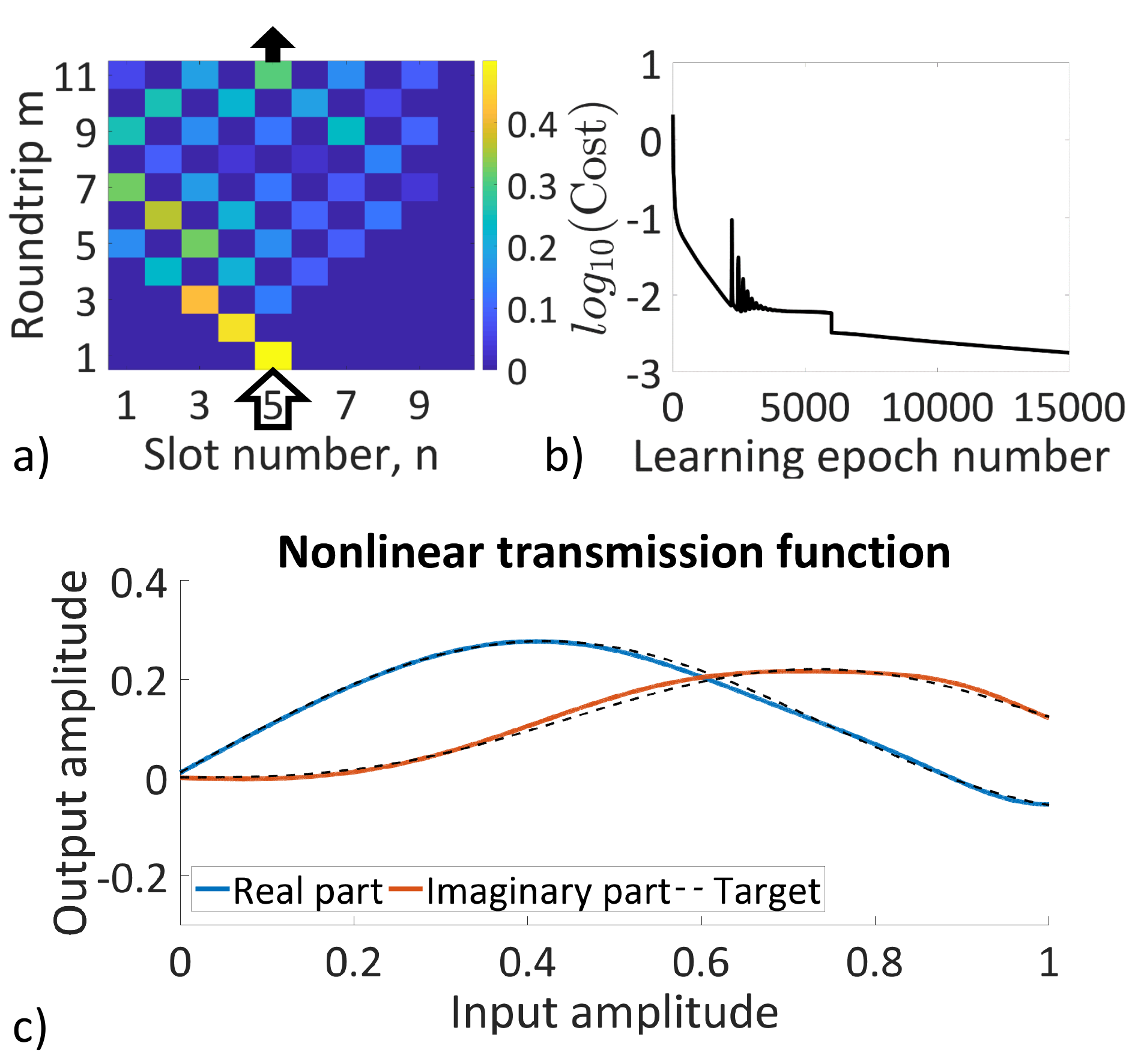}
\caption{A synthetic photonic lattice performing nonlinear transformation combining nonlinear saturation and phase modulation, $u_5^{11}=u_5^0\exp(2 (i-1) |u_5^0|^2)$. (a)~Signal evolution in the short loop of the SPL, color shading represents the signal intensity. An input signal is launched into a single slot $n=5$, the output signal is read from the same slot after 11 roundtrips. (b)~Cost function vs. epoch. (c)~Real and imaginary parts of the output vs. the input amplitude for the trained lattice (solid lines) compared with the target nonlinear function (dashed lines).}
\label{fig:transmission}
\end{figure}

As a representative case, we consider the central slot $u_5^0$ of the nine slot SPL as an input and require the output signal after 11 roundtrips to be $u_5^{11}=u_5^0\exp(2 (i-1) |u_5^0|^2)$, realising both saturation and nonlinear phase modulation effects. An example of signal evolution in the network is shown in Fig.~\ref{fig:transmission}(a). 
Optimization of the network parameters is performed using sets of different input signal amplitudes. As expected, a larger number of learning epochs are required compared to a linear system, 
see the evolution of cost function during the training process, Fig.~\ref{fig:transmission}(b).  We used only 10 sets for training the ONN, which represented 10 different amplitudes of input signal, as this allowed for the best convergence. Most importantly, the final nonlinear transmission reproduces very closely the target function, as checked for $10^4$ input amplitude values in the selected range shown in Fig.~\ref{fig:transmission}(c). Note that the output is accurately matched in terms of both the intensity and phase. This demonstrates the capacity of SPL-based ONN to perform complex nonlinear transformations.

\subsection{Inference training for the discrimination of input states}
\label{subsec:vector_separation}
Finally, we reveal the capabilities of the nonlinear network for state discrimination, targeting the operation where different inputs from a defined set are mapped to distinct single-site outputs. We consider a nontrivial case when the input states, defined as complex vectors, are non-orthogonal. On the other hand, we aim to have fully orthogonal outputs. Such task of non-orthogonal to orthogonal mapping can be realized by a specially optimized non-conservative linear transformation only for a set of two complex input vectors~\cite{Cerjan:2017-253902:PRL, Lung:2020-3015:ACSP}. However, for a set with three or more input elements, nonlinear transformation is required for complete discrimination, and we illustrate how this can be accomplished by the SPL training for inference.

\begin{figure}[ht]
\centering
\includegraphics[width=\linewidth]{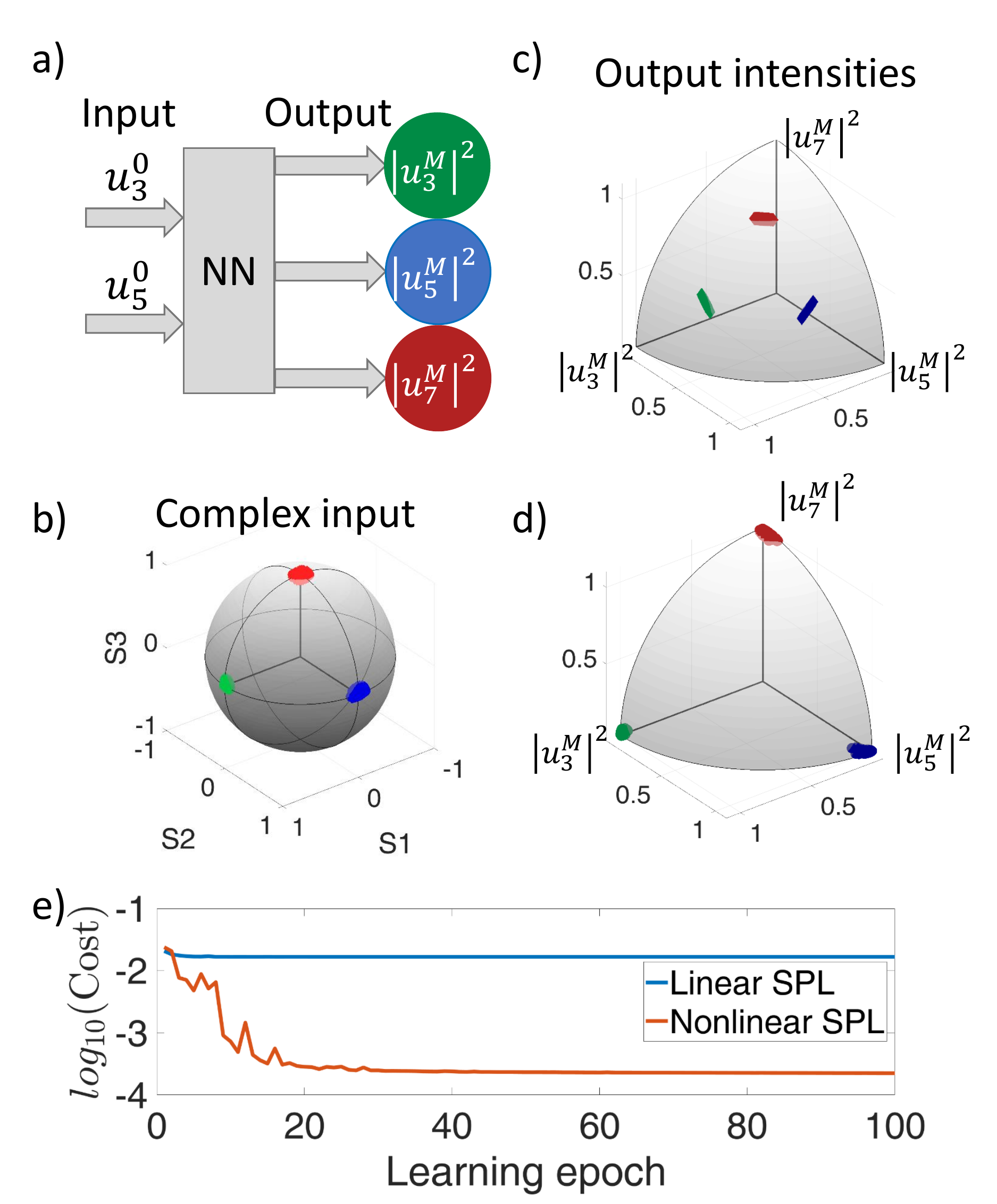}
\caption{Discrimination of input states with an SPL. (a)~Schematic diagram of target performance, where each of the three input vector states coupled to slots 3 and 5 is mapped to one of the selected output slots. (b)~Sets of input states represented on the Poincare sphere. (c,d)~The output intensities $(|u_3^M|^2, |u_5^M|^2,|u_7^M|^2)$ for the trained (c)~linear and (d)~nonlinear SPL networks. The colors indicate different target outputs as marked in~(a). (e)~Cost function vs. training epoch for linear and nonlinear SPLs.}
\label{fig:vector_clouds}
\end{figure}

We demonstrate the discrimination between a set of three non-orthogonal complex vectors represented by complex signal amplitudes $[u_3^0, u_5^0]$ at the third and fifth input slots of the SPL with the dimensions $(M,N)=(11,5)$. We then demand the output signal of the synthetic photonic lattice to concentrate in one of the selected output slots $|u_3^M|^2, |u_5^M|^2,|u_7^M|^2$ depending on the input, see Fig.~\ref{fig:vector_clouds}(a). As an example, we illustrate the training for the input states corresponding to the poles on the Poincare sphere representation, see Fig.~\ref{fig:vector_clouds}(b). We prepare the training sets to include small deviations at the inputs, shown as circular regions on the Poincare sphere.


For comparison, we perform the training of linear ($\alpha=0$) and nonlinear ($\alpha=1$) SPLs and show the intensities at the three target output slots for different inputs in Figs.~\ref{fig:vector_clouds}(c) and~(d), respectively.
We observe that a nonlinear ONN achieves near-perfect discrimination with the outputs concentrated at only the selected slots for different inputs [Fig.~\ref{fig:vector_clouds}(d)], whereas the outputs spread out across all slots indicating a fundamentally restricted discrimination performance of a linear network [Fig.~\ref{fig:vector_clouds}(c)]. The advantage of a nonlinear network is confirmed by comparison of the cost function evolution during the training process, reaching two orders of magnitude lower values compared to a linear network, see Fig.~\ref{fig:vector_clouds}(e). 
We have furthermore confirmed the scalability of a nonlinear SPL for discrimination between a larger number of input states, see Appendix~\ref{appendix}.

\section{Conclusion}
We demonstrated that synthetic photonic lattices operating with optical pulse sequences can function as artificial feed-forward neural networks. We also showed that Kerr nonlinearity within the lattice fundamentally enhances the optical network capabilities beyond the constraints of linear transformations.
%
Furthermore, the design flexibility of synthetic photonic lattice structures may pave the way towards multidimensional ONN operating in a synthetic space~\cite{Schreiber:2012-55:SCI, Muniz:2019-9518:SRP, Muniz:2019-253903:PRL}.
We anticipate that our results can stimulate the experimental development of synthetic photonic lattices for complex real-time processing of optical pulses based on the principles of artificial deep neural networks.

\begin{acknowledgments}
This work is supported by Ministry of Education and Science of the Russian Federation (FSUS-2020-0034) and the Australian Research Council (DP190100277). We thank Dr. Oleg Sidelnikov for valuable discussions and comments. The simulation data underlying the results presented in this paper may be obtained from the authors upon reasonable request.
\end{acknowledgments}

\appendix
\section{Discrimination of multiple input states}
\label{appendix}

\begin{figure}[th]
\centering
\includegraphics[width=\linewidth]{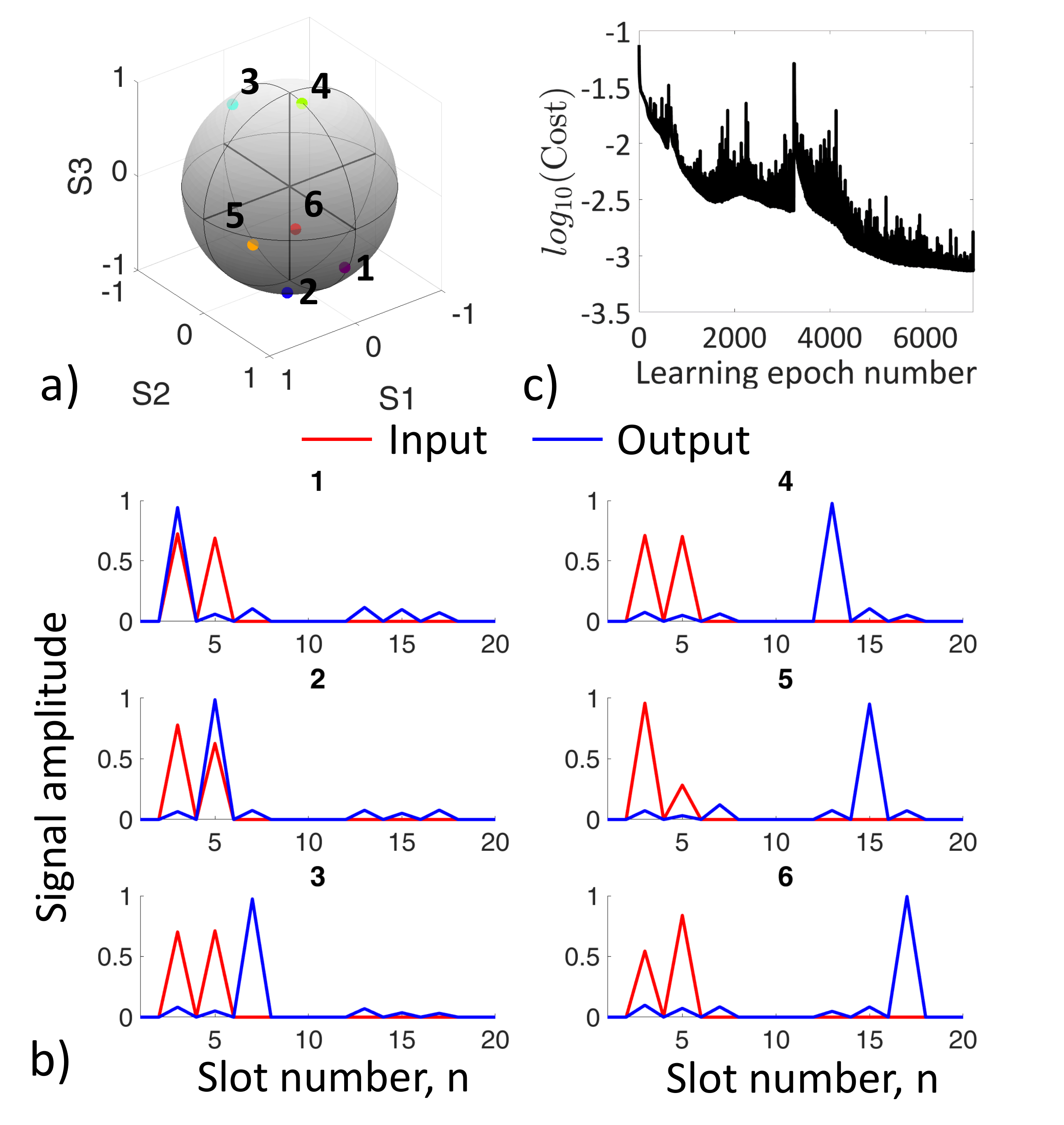}
\caption{Discrimination between 6 input states by a nonlinear SPL.
(a)~Sets of the input states represented on Poincare sphere. 
(b)~ Intensities at the input and output of a trained SPL averaged over 100 realisations of different states as indicated by labels. (c)~Cost function vs. the learning epoch.}
\label{fig:6_vectors}
\end{figure}

We demonstrate that a nonlinear SPL can be trained to discriminate between 6 different inputs, confirming that the operation shown in Sec.~\ref{subsec:vector_separation} is scalable to a larger number of states. Specifically, we aim at mapping 6 different complex vectors [Fig.~\ref{fig:6_vectors}(a)] from two input slots to 6 separate outputs of the ONN.
We tested the operation for bunches of 100 input vectors in the vicinity of the 6 input states, and confirmed that the trained SPL demonstrates excellent capability for state discrimination, see Fig.~\ref{fig:6_vectors}(b). This is confirmed by a good convergence of the cost function as illustrated in Fig.~\ref{fig:6_vectors}(c).

\bibliography{db_art_neural_SPL}

\end{document}